\documentclass[12pt,a4paper]{article}

\newcommand{\la}{\lambda}
\newcommand{\pa}{\partial}
\newcommand{\oa}{{\omega}_1}
\newcommand{\ob}{{\omega}_2}
\newcommand{\al}{\alpha}

\begin{document}

\begin{flushright}
{}
\end{flushright}
\vspace{1.8cm}

\begin{center}
 \textbf{\Large Giant Magnon and Spike Solutions \\
with Two Spins in $AdS_4 \times CP^3$ }
\end{center}
\vspace{1.6cm}
\begin{center}
 Shijong Ryang
\end{center}

\begin{center}
\textit{Department of Physics \\ Kyoto Prefectural University of Medicine
\\ Taishogun, Kyoto 603-8334 Japan}  \par
\texttt{ryang@koto.kpu-m.ac.jp}
\end{center}
\vspace{2.8cm}
\begin{abstract}
In the string theory in $AdS_4 \times CP^3$ we construct the giant 
magnon and spike solutions with two spins in two kinds of subspaces of
$R_t \times CP^3$ and derive the dispersion relations for them. For 
the single giant magnon solution in one subspace we show that its
dispersion relation is associated with that of the big one-spin giant
magnon solution in the $RP^2$ subspace. For the single giant magnon
solution in the other complementary subspace its dispersion relation
is similar to that of the one-spin giant magnon solution living in 
the $S^2$ subspace but has one additional spin dependence. 
\end{abstract} 
\vspace{3cm}
\begin{flushleft}
September, 2008
\end{flushleft}

\newpage
\section{Introduction}

There has been an exciting subject toward understanding the correspondence
 between the four-dimensional $\mathcal{N}=4$  super Yang-Mills
theory and the type IIB superstring in $AdS_5 \times S^5$ \cite{MGW}.
Recently, inspired by the study of Bagger, Lambert and Gustavsson about
the three-dimensional  $\mathcal{N}=8$ supersymmetric theory for the
multiple M2-branes \cite{BLG}, Aharony, Bergman, Jafferis and Maldacena
(ABJM) have constructed a $\mathcal{N}=6$ superconformal $SU(N) \times
SU(N)$ Chern-Simons theory in three dimensions at level $(k,-k)$ coupled
with bi-fundamental matter as a world-volume theory of $N$ coincident
M2-branes on orbifold $C^4/Z_k$ \cite{ABJM}.
When the 't Hooft coupling constant
defined by $\la =N/k$ becomes large and $1 \ll
N \ll k^5$, the M-theory with $N$-units of four-form flux on
$AdS_4 \times S^7/Z_k$ can be compactified to the type IIA superstring
theory in $AdS_4 \times CP^3$. 

The new correspondence between the ABJM theory and  the type IIA 
superstring theory in $AdS_4 \times CP^3$ has been studied by analyzing
the string spectrum in the Penrose limit of this IIA background
\cite{NT,GHO,GGY} and by finding an integrable Hamiltonian of an $SU(4)$
spin chain \cite{MZ,GGY} with sites alternating between the fundamental
and anti-fundamental representations and further constructing two-loop 
Bethe equation \cite{MZ,BR}.  The all-loop 
Bethe ansatz equation has been obtained
\cite{GV} by developing the perturbative result \cite{MZ} and
the classical integrability in the strong coupling which was shown by
deriving a supercoset sigma model for the type IIA superstring theory in
$AdS_4 \times CP^3$ \cite{AF} and an $OSp(2,2|6)$ symmetric finite gap
algebraic curve \cite{NGV}. This all-loop Bethe ansatz equation has 
been also produced by analyzing the S-matrix \cite{AN}.

There have been a construction of a giant magnon solution with one angular
momentum in the $SU(2) \times SU(2)$ sector of type IIA string theory in
$AdS_4 \times CP^3$ \cite{GHO} (see also \cite{GGY}) and a computation of
its finite-size correction \cite{GHOS}, where the string moving in
$R_t \times S^2 \times S^2$ has opposite azimuthal angles in the two 
$S^2$. The dispersion relation of the spike solution with one spin
 in $R_t \times S^2 \times S^2$ and its finite-size correction have
been derived \cite{LPP}. The circular, folded and pulsating string 
solutions in the $SU(2) \times SU(2)$ sector have been studied \cite{CW}.
Further, the giant magnon and spike solutions with two angular momenta 
have been constructed by reducing the string dynamics on 
$AdS_4 \times CP^3$ to the Neumann-Rosochatius integrable system 
\cite{ABR}, where the dispersion relations and the finite-size corrections
for them have been computed. 
On the other hand from the analysis of the finite gap algebraic curve the
dispersion relation of the giant magnon and the one-loop quantum 
correction have been presented \cite{IS}. The $AdS_4 \times CFT_3$ duality
has been further investigated from various view points 
\cite{MR,AAB,CK,GM,BGR,BT,RR,MRT}.

In ref. \cite{ABR} the four embedding complex coordinates constrained by a
$CP^3$ condition have been used for parametrizing the $CP^3$ space to 
construct the string moving in $R_t \times S^3 \times S^3$ with two
angular momenta in the first $S^3$ and with exactly opposite angular
momenta in the second $S^3$, and extract the dispersion relation of a 
single giant magnon solution living in $S^3$ with two angular momenta.
We will choose the spherical coordinates of $CP^3$ space presented
in ref. \cite{CLP,NT} to construct a 
single giant magnon solution as well as
a single spike solution with two angular momenta living in one subspace of
$CP^3$ which is different from the $S^3$ subspace and derive
the dispersion relations for them. By considering the other complementary
subspace of $CP^3$ we will obtain some different types of giant magnon and
spike solutions with two angular momenta.

\section{Giant magnon and spike solutions in one subspace of $CP^3$}

Following the prescriptions to construct the giant magnon solutions 
\cite{HM,AFZ,AFGS} and the spike solutions \cite{IK,NBR} in 
$AdS_5 \times S^5$, we consider the string 
states such as a giant magnon and
a spike in $AdS_4 \times CP^3$ which have two angular momenta in $CP^3$.
The explicit form of metric on $AdS_4 \times CP^3$ is 
written by \cite{CLP,NT}
\begin{eqnarray}
ds^2 &=& \frac{R^2}{4}[-\cosh^2\rho dt^2 + d\rho^2 + \sinh^2\rho
d\Omega_2^2 ] \nonumber \\
&+& R^2\biggl[d\xi^2 + \cos^2\xi\sin^2\xi
\left(d\psi + \frac{1}{2}\cos\theta_1
d\phi_1 -  \frac{1}{2}\cos\theta_2 d\phi_2\right)^2 \nonumber \\
&+& \frac{1}{4}\cos^2\xi(d\theta_1^2 + \sin^2\theta_1d\phi_1^2)
+ \frac{1}{4}\sin^2\xi(d\theta_2^2 + \sin^2\theta_2d\phi_2^2) \biggr],
\end{eqnarray}
where $0 \le \xi \le \pi/2, -2\pi \le \psi \le 2\pi$ and 
$(\theta_i,\phi_i)$ are coordinates of two $S^2$'s. 
The radius $R$ is given by 
$R^2 = 2^{5/2}\pi \la^{1/2}$ with the 't Hooft coupling constant 
$\la = N/k$. We are interested in a 
string configuration in $R_t \times CP^3$
which is parametrized by $\rho = 0$ as well as $\theta_1 = \theta_2
\equiv \theta$. 

Here with this parametrization we write down the Polyakov
action in the conformal gauge 
\begin{eqnarray}
S &=& \sqrt{2\la}\int d\tau d\sigma  \biggl[ 
-\frac{1}{4}\dot{t}^2 + \dot{\xi}^2
- {\xi'}^2 + \cos^2\xi\sin^2\xi( \dot{\psi}^2 - {\psi'}^2 ) \nonumber \\
&+& \frac{1}{4}( \dot{\theta}^2 - {\theta'}^2 ) + \cos^2\xi \sin^2\xi
\cos\theta \left( \dot{\psi}(\dot{\phi_1} - \dot{\phi_2} ) -
\psi'( \phi'_1 - \phi'_2 ) \right) \nonumber \\
&+& \frac{1}{4}\cos^2\xi ( \sin^2\theta + \cos^2\theta\sin^2\xi)
( \dot{\phi_1}^2 - {\phi'_1}^2 ) + \frac{1}{4}\sin^2\xi ( \sin^2\theta + 
\cos^2\theta\cos^2\xi)( \dot{\phi_2}^2 - {\phi'_2}^2 ) \nonumber \\
&-& \frac{1}{2}\cos^2\xi \sin^2\xi \cos^2\theta( \dot{\phi_1}\dot{\phi_2}
-  \phi'_1\phi'_2 ) \biggr].
\end{eqnarray}
The equation of motion for $\theta$ is given by
\begin{eqnarray}
\ddot{\theta} - \theta'' - \sin\theta \cos\theta [ (\cos^2\xi \dot{\phi_1}
+ \sin^2\xi \dot{\phi_2})^2 - (\cos^2\xi \phi'_1 + \sin^2\xi \phi'_2)^2 ]
\nonumber \\
+ 2\cos^2\xi \sin^2\xi \sin\theta [ \dot{\psi}(\dot{\phi_1} 
- \dot{\phi_2} ) - \psi'( \phi'_1 - \phi'_2 ) ] = 0,
\label{the}\end{eqnarray}
while the equation of motion for $\psi$ reads
\begin{eqnarray}
2[\pa_{\tau}(\cos^2\xi\sin^2\xi \dot{\psi}) - \pa_{\sigma}
(\cos^2\xi\sin^2\xi \psi') ] + \pa_{\tau}[\cos^2\xi \sin^2\xi \cos\theta
( \dot{\phi_1} - \dot{\phi_2} )] \nonumber \\
 -  \pa_{\sigma}[\cos^2\xi \sin^2\xi \cos\theta ( \phi'_1
 - \phi'_2 )] = 0.
\label{pse}\end{eqnarray}
From the two equations we note that there is an obvious solution
\begin{equation}
\theta = \frac{\pi}{2}, \hspace{1cm} \psi = 0.
\label{obv}\end{equation}
Then the relevant string action is
\begin{equation}
S = \sqrt{2\la}\int d\tau d\sigma  \left[ -\frac{1}{4}\dot{t}^2 + 
\dot{\xi}^2 - {\xi'}^2 
+ \frac{1}{4}\cos^2\xi ( \dot{\phi_1}^2 - {\phi'_1}^2 ) 
+ \frac{1}{4}\sin^2\xi ( \dot{\phi_2}^2 - {\phi'_2}^2 ) \right],
\end{equation}
where $\tau$ and $\sigma$ range from $-\infty$ to $\infty$. 

Let us make the following ansatz for a rotating string soliton
with two spins
\begin{equation}
t = \kappa\tau, \hspace{1cm} \xi = \xi(\eta), \hspace{1cm}  
 \phi_1 = \oa\tau + f_1(\eta), \hspace{1cm} 
\phi_2 = \ob\tau + f_2(\eta), 
\end{equation}
where $\eta = \al \sigma + \beta \tau$. The equations of motion for
$\phi_1, \phi_2$ lead to
\begin{equation}
\pa_{\eta}f_1 = \frac{1}{\al^2 - \beta^2}\left( \frac{C_1}{\cos^2\xi}
+ \beta \oa \right), \hspace{1cm} \pa_{\eta}f_2 = \frac{1}
{\al^2 - \beta^2}\left( \frac{C_2}{\sin^2\xi} + \beta \ob \right),
\end{equation}
where $C_1$ and $C_2$ are integration constants. The Virasoro constraints
$T_{\tau \tau} + T_{\sigma \sigma} = 0, T_{\tau \sigma} = 0$ yield
\begin{eqnarray}
- \frac{\kappa^2}{4(\al^2 + \beta^2)} + {\xi'}^2 + \frac{\cos^2\xi}{4}
\left( {f_1'}^2 + \frac{\oa^2 + 2\oa \beta f_1'}{\al^2 + \beta^2} \right)
+ \frac{\sin^2\xi}{4}\left( {f_2'}^2 + \frac{\ob^2 + 2\ob \beta f_2'}
{\al^2 + \beta^2} \right) = 0, \nonumber \\
{\xi'}^2 + \frac{\cos^2\xi}{4}\left( {f_1'}^2 + \frac{\oa f_1'}{\beta} 
\right) + \frac{\sin^2\xi}{4}\left( {f_2'}^2 + 
\frac{\ob f_2'}{\beta} \right) = 0,
\end{eqnarray}
where the prime here denotes the derivative with respect to $\eta$.
The difference of these two constraints gives a relation between few
parameters
\begin{equation}
\beta \kappa^2 + \oa C_1 + \ob C_2 = 0.
\end{equation}

We restrict ourselves to $C_1 = 0$ for constructing a giant magnon 
solution and a spike solution so that $C_2$ is given by
\begin{equation}
C_2 = - \frac{\beta\kappa^2}{\ob}.
\label{ctw}\end{equation}
Combining together we have
\begin{equation}
{\xi'}^2 = \frac{1}{4(\al^2 - \beta^2)^2} \left[ \kappa^2(\al^2 + \beta^2)
- \frac{C_2^2}{\sin^2\xi} - \al^2( \oa^2\cos^2\xi + \ob^2\sin^2\xi )
 \right],
\end{equation}
which is the first integral of the equation of motion for $\xi$.
For $\ob > \oa$ this equation is expressed in the following form
\begin{equation}
\xi' = \pm \frac{\al\sqrt{\ob^2 - \oa^2}}{2(\al^2 - \beta^2)\sin\xi} 
\sqrt{(\cos^2\xi_+ - \cos^2\xi)(\cos^2\xi - \cos^2\xi_-)},
\end{equation}
where the $+$ and $-$ signs correspond to $\al^2 > \beta^2$ and  
$\al^2 < \beta^2$ respectively and
\begin{eqnarray}
\cos^2\xi_{\pm} &=& \frac{1}{2}\biggl[ \frac{\al^2(2\ob^2 - \oa^2)
- \kappa^2(\al^2 +\beta^2)}{\al^2(\ob^2 - \oa^2)} \nonumber \\
&\pm& \sqrt{ \left( \frac{\al^2(2\ob^2 - \oa^2)
 - \kappa^2(\al^2 +\beta^2)}
{\al^2(\ob^2 - \oa^2)} \right)^2 - 4 \frac{ \al^2\ob^2 + C_2^2 
-\kappa^2(\al^2 + \beta^2) }{ \al^2(\ob^2 - \oa^2)}  }  \biggr]
\label{cor}\end{eqnarray}
with $\cos\xi_+ > \cos\xi_-$. 

The string energy is provided by
\begin{equation}
E = \pm 2\sqrt{2\la} \frac{\kappa(\al^2 - \beta^2)}{\al^2 
\sqrt{\ob^2 - \oa^2}} \int_{\xi_+}^{\xi_-}d\xi \frac{\sin\xi}
{\sqrt{(\cos^2\xi_+ - \cos^2\xi)(\cos^2\xi - \cos^2\xi_-)} },
\end{equation}
while the spins $J_1$ and $J_2$ associated with the angular variables
$\phi_1$ and $\phi_2$ are expressed as
\begin{eqnarray}
J_1 &=& \pm 2\sqrt{2\la} \frac{\al^2 - \beta^2}{\al^2 
\sqrt{\ob^2 - \oa^2}} \int_{\xi_+}^{\xi_-}d\xi \frac{\sin\xi\cos^2\xi}
{\sqrt{(\cos^2\xi_+ - \cos^2\xi)(\cos^2\xi - \cos^2\xi_-)} }
\left( \oa + \frac{\beta^2\oa}{\al^2 - \beta^2} \right), \nonumber \\
J_2 &=& \pm 2\sqrt{2\la} \frac{\al^2 - \beta^2}{\al^2 
\sqrt{\ob^2 - \oa^2}} \int_{\xi_+}^{\xi_-}d\xi \frac{\sin^3\xi}
{\sqrt{(\cos^2\xi_+ - \cos^2\xi)(\cos^2\xi - \cos^2\xi_-)} } 
\nonumber \\
&\times& \left[ \ob + \frac{\beta}{\al^2 - \beta^2}
\left( \frac{C_2}{\sin^2\xi} + \beta \ob \right) \right].
\end{eqnarray}
The angle difference for $\phi_2$ is also described by
\begin{eqnarray}
\Delta \phi_2 &=& - \int d\phi_2 = - \int d\xi \frac{\phi_2'}
{\xi'} \nonumber \\
&=& \mp \frac{4}{\al \sqrt{\ob^2 - \oa^2}} \int_{\xi_+}^{\xi_-}d\xi 
\frac{\sin\xi}{\sqrt{(\cos^2\xi_+ - \cos^2\xi)(\cos^2\xi - \cos^2\xi_-)} }
\left( \frac{C_2}{\sin^2\xi} + \beta \ob \right),
\label{ang}\end{eqnarray}
where we use a minus sign to make the angle difference positive.
The giant magnon and spike solutions in the infinite volume can be 
constructed by taking a limit $\xi_- \rightarrow \pi/2$.
Therefore the infinite volume limit specified by $\cos^2\xi_- = 0$ in
(\ref{cor}) yields a relation 
\begin{equation}
\kappa^2 ( \al^2 + \beta^2 ) = \al^2 \ob^2 + C_2^2,
\end{equation}
which becomes the following equation through (\ref{ctw})
\begin{equation}
(\kappa^2 - \ob^2 )\left( \al^2 - \frac{\kappa^2\beta^2}{\ob^2} 
\right) = 0.
\end{equation}
There are two solutions $\kappa = \ob$ and $\kappa = \al \ob/\beta$,
which are associated with the giant magnon 
and spike solutions respectively.
From (\ref{cor}), $\sin^2\xi_+$ for each solution is evaluated as
\begin{equation}
\sin^2\xi_+ = \frac{\ob^2\beta^2}{(\ob^2 - \oa^2)\al^2}
\;\; (\mathrm{giant\ magnon}), \hspace{1cm}
\sin^2\xi_+ = \frac{\ob^2\al^2}{(\ob^2 - \oa^2)\beta^2}
\; \;(\mathrm{spike})
\end{equation}
so that $\al^2 > \beta^2$ and $\al^2 < \beta^2$ correspond to the
giant magnon and spike solutions respectively.

For the giant magnon solution case the difference between the infinite 
energy $E$ and the infinite spin $J_2$ becomes finite as
\begin{equation}
E - J_2 = 2\sqrt{2\la} \frac{\ob \cos\xi_+}{\sqrt{ \ob^2 - \oa^2} },
\end{equation}
which is also expressed as $E - J_2 = J_1 \ob/\oa$ in terms of 
the finite spin $J_1$. The elimination of $\ob/\oa$ leads to
\begin{equation}
E - J_2 = \sqrt{J_1^2 + 8\la \cos^2\xi_+ }.
\end{equation}
Since $C_2/\sin^2\xi + \beta \ob$ in (\ref{ang}) becomes 
$-\beta\ob \cos^2\xi/\sin^2\xi$, the angle difference is given by 
a positive finite value
\begin{eqnarray}
\Delta \phi_2 &=& 4 \sin\xi_+ \int_{\xi_+}^{\pi/2} d\xi 
\frac{\cos\xi}{\sin\xi \sqrt{\sin^2\xi - \sin^2\xi_+} } \nonumber \\
&=& 4 \left( \frac{\pi}{2} - \xi_+ \right).
\end{eqnarray}
The magnon momentum $p$ is identified with the angle difference 
$\Delta \phi_2$ so that we get a dispersion relation of a single giant
magnon in one subspace of $R_t \times CP^3$
\begin{equation}
E - J_2 = \sqrt{J_1^2 + 8\la \sin^2\frac{p}{4} },
\label{dis}\end{equation}
which is of a similar form to that of the dyonic giant magnon in
$AdS_5 \times S^5$ \cite{ND}. For comparison we write down the dispersion
relation of a single two-spin giant magnon living in the $S^3$ subspace
\begin{equation}
E - J_2 = \sqrt{\frac{J_1^2}{4} + 2\la \sin^2\frac{p}{2} },
\label{ejd}\end{equation}
which was extracted \cite{ABR} by analyzing the string motion in 
$R_t \times S^3 \times S^3$ as the Neumann-Rosochatius integrable system.
When $J_1$ is turned off the dispersion relation (\ref{ejd}) reduces to
that of a single one-spin giant magnon 
living in the $S^2$ subspace \cite{GHO}.
The result (\ref{dis}) with a dependence of $\sin^2p/4$ shows the same
expression as the dispersion relation for a single big giant magnon
solution with two spins in $AdS_4 \times CP^3$ in ref. \cite{IS}
(see \cite{GGY}), where it was derived by using the algebraic curve 
technique \cite{BKSZ,KMMZ}, and the other dispersion relation (\ref{ejd})
was also computed separately. 

Let us consider the single spike solution case. Since $C_2/\sin^2\xi +
\beta\ob$ in (\ref{ang}) is expressed by 
$\ob(\beta^2\sin^2\xi - \al^2)/\beta\sin^2\xi$ which is positive for
$\pi/2 > \xi > \xi_+$ owing to $\beta^2\sin^2\xi_+ - \al^2 = \al^2
\oa^2/(\ob^2 - \oa^2) > 0$. The angle difference $\Delta \phi_2$ becomes
positive divergent. The difference between the infinite energy $E$ and
the infinite positive angle difference $\Delta \phi_2$ multiplied with
$\sqrt{2\la}/2$ turns out to be finite
\begin{eqnarray}
E - \frac{1}{2}\sqrt{2\la}\Delta \phi_2 &=& 2\sqrt{2\la} \sin\xi_+ 
\int_{\xi_+}^{\pi/2} d\xi \frac{\cos\xi}
{\sin\xi \sqrt{\sin^2\xi - \sin^2\xi_+} } \nonumber \\
&=& 2\sqrt{2\la}\bar{\xi}
\label{hah}\end{eqnarray}
with $\bar{\xi} = \pi/2 - \xi_+$. On the other hand the two spins 
$J_1$ and $J_2$ are finite as expressed by
\begin{eqnarray}
J_1 &=& - 2\sqrt{2\la} \frac{\oa \cos\xi_+}{\sqrt{ \ob^2 - \oa^2} },
\nonumber \\
J_2 &=& 2\sqrt{2\la} \frac{\ob \cos\xi_+}{\sqrt{ \ob^2 - \oa^2} },
\end{eqnarray}
which satisfy a relation
\begin{equation}
J_2 = \sqrt{J_1^2 + 8\la \sin^2\bar{\xi} }.
\end{equation}

\section{Giant magnon and spike solutions in the complementary subspace
of $CP^3$}

We construct the other string solution by making the following ansatz
\begin{eqnarray}
t &=& \kappa \tau, \hspace{1cm} \theta = \frac{\pi}{2}, \hspace{1cm}
\psi = \omega \tau + f(\eta), \nonumber \\
\xi &=& \xi(\eta), \hspace{1cm} \phi_1 = \phi_2 \equiv \phi
\end{eqnarray} 
with $\phi = \nu\tau$.
The equation of motion for $\theta$ (\ref{the})
 is solved by@$\theta = \pi/2$
and $\phi_1 = \phi_2$, which is a complementary choice to 
(\ref{obv}). The equation of motion for $\phi$ is simply satisfied,
while the equation of motion for $\psi$ (\ref{pse}) yields 
\begin{equation}
f' = \frac{1}{\al^2 - \beta^2} \left( \frac{C}{\sin^22\xi}
+ \beta\omega \right)
\end{equation}
with an integration constant $C$. The Virasoro constraints read
\begin{eqnarray}
- \frac{\kappa^2}{4(\al^2 + \beta^2)} + {\xi'}^2
+ \frac{\sin^22\xi}{4}\left( {f'}^2 + \frac{\omega^2 + 2\omega \beta f'}
{\al^2 + \beta^2} \right) + \frac{\nu^2}{4(\al^2 + \beta^2)}
= 0, \nonumber \\
{\xi'}^2 + \frac{\sin^22\xi}{4}\left( {f'}^2 + 
\frac{\omega f'}{\beta} \right) = 0,
\end{eqnarray} 
whose difference provides a relation $\beta (\kappa^2 - \nu^2) 
+ \omega C = 0$. We gather them together to have
\begin{equation}
{\xi'}^2 = \frac{1}{4(\al^2 - \beta^2)^2} \left[ (\kappa^2 - \nu^2 )
(\al^2 + \beta^2) - \frac{C^2}{\sin^22\xi} 
- \al^2\omega^2\sin^22\xi  \right],
\end{equation}
which yields
\begin{equation}
\xi' = \pm \frac{\al\omega}{2(\al^2 - \beta^2)\sin2\xi} 
\sqrt{(\cos^22\xi_+ - \cos^22\xi)(\cos^22\xi - \cos^22\xi_-)},
\end{equation}
where the $+$ and $-$ signs correspond to the giant magnon 
($\al^2 > \beta^2$) and the spike ($\al^2 < \beta^2$) solutions and
\begin{eqnarray}
\cos^22\xi_{\pm} &=& \frac{1}{2}\biggl[ \frac{2\al^2\omega^2 - 
(\kappa^2 - \nu^2)(\al^2 + \beta^2)}{\al^2\omega^2}  \label{cog}  \\
&\pm& \sqrt{ \left(  \frac{2\al^2\omega^2 - 
(\kappa^2 - \nu^2)(\al^2 + \beta^2)}{\al^2\omega^2} \right)^2 
- 4 \frac{ \al^2\omega^2 + C^2 -(\kappa^2 - \nu^2)(\al^2 + \beta^2) }
{ \al^2\omega^2 } } \biggr]. \nonumber
\end{eqnarray}

The rotating string is specified by the following energy and two spins
$J_{\phi}$ and $J_{\psi}$ associated with the $\phi$ and $\psi$ directions
\begin{eqnarray}
E &=& \pm 2\sqrt{2\la} \frac{\kappa(\al^2 - \beta^2)}{\al^2 
\omega} \int_{\xi_+}^{\xi_-}d\xi \frac{\sin2\xi}{\sqrt{(\cos^22\xi_+ 
- \cos^22\xi)(\cos^22\xi - \cos^22\xi_-)} }, \nonumber \\
J_{\phi} &=& \pm 2\sqrt{2\la} \frac{\nu(\al^2 - \beta^2)}{\al^2 
\omega} \int_{\xi_+}^{\xi_-}d\xi \frac{\sin2\xi}{\sqrt{(\cos^22\xi_+ 
- \cos^22\xi)(\cos^22\xi - \cos^22\xi_-)} }, \nonumber \\
J_{\psi} &=& \pm 2\sqrt{2\la} \frac{\al^2 - \beta^2}{\al^2 
\omega} \int_{\xi_+}^{\xi_-}d\xi \frac{\sin^32\xi}
{\sqrt{(\cos^22\xi_+ - \cos^22\xi)(\cos^22\xi - \cos^22\xi_-)} } 
\nonumber \\
 &\times& \left[ \omega + \frac{\beta}{\al^2 - \beta^2}
\left( \frac{C}{\sin^22\xi} + \beta \omega \right) \right].
\label{ejp}\end{eqnarray}
The angle difference for $\psi$ is given by
\begin{eqnarray}
\Delta \psi &=& - \int d\psi \nonumber \\
&=& \mp \frac{4}{\al \omega}\int_{\xi_+}^{\xi_-} d\xi  \frac{\sin2\xi}
{\sqrt{(\cos^22\xi_+ - \cos^22\xi)(\cos^22\xi - \cos^22\xi_-)} } 
\left( \frac{C}{\sin^22\xi} + \beta \omega \right).
\label{dep}\end{eqnarray}
Here we take an infinite volume limit $\xi_- \rightarrow \pi/4$ 
in (\ref{cog}) to obtain
an equation
\begin{equation}
(\kappa^2 - \nu^2 - \omega^2 )\left( \al^2 - \frac{\beta^2
(\kappa^2 - \nu^2)}{\omega^2} \right) = 0,
\end{equation}
whose solutions $\kappa^2 - \nu^2 = \omega^2$ and $\kappa^2 - \nu^2 = 
\al^2\omega^2/\beta^2$ specify the giant magnon and spike solutions
in the infinite volume respectively. From (\ref{cog}) each solution is
characterized by
\begin{equation}
\sin^22\xi_+ = \frac{\beta^2}{\al^2} \; \;(\mathrm{giant\ magnon}),
\hspace{1cm} \sin^22\xi_+ = \frac{\al^2}{\beta^2} \; \;(\mathrm{spike}),
\end{equation}
which implies that $\al^2 > \beta^2$ and $\al^2 < \beta^2$ are associated
with the giant magnon and spike solutions respectively.

We begin to consider the giant magnon solution case. In view of the 
expressions in (\ref{ejp}) we find  one relation between the 
energy $E$ and the spins $J_{\phi}$ and $J_{\psi}$ 
in the infinite volume
\begin{eqnarray}
\kappa E - \nu J_{\phi} - \omega J_{\psi} &=& 2\sqrt{2\la} \omega
\int_{\xi_+}^{\pi/4}d\xi \frac{\sin2\xi\cos2\xi}{\sqrt{\cos^22\xi_+ 
- \cos^22\xi}} \nonumber \\ 
&=& \sqrt{2\la}\omega \cos2\xi_+,
\label{rel}\end{eqnarray}
where the logarithmic divergences of $E, J_{\phi}$ and $J_{\psi}$
are canceled out under the suitably chosen coefficients.
There is the other relation
\begin{equation}
\frac{E}{\kappa} = \frac{J_{\phi}}{\nu}.
\label{lin}\end{equation}
From (\ref{dep}) the angle difference identified with the magnon momentum
$p$ is given by
\begin{equation}
p = \Delta \psi = 2\left( \frac{\pi}{2} - 2\xi_+ \right).
\label{pan}\end{equation}
Combining (\ref{rel}), (\ref{lin}) and (\ref{pan}) with 
\begin{equation}
\left(\frac{\kappa}{\omega}\right)^2 - \left(\frac{\nu}{\omega}
\right)^2 = 1
\end{equation}
to eliminate the parameters we get the following  
dispersion relation for this rotating two-spin string solution
\begin{equation}
\sqrt{E^2 - J_{\phi}^2} - J_{\psi} = \sqrt{2\la}\sin\frac{p}{2}.
\label{sqe}\end{equation}
In the $J_{\phi}=0$ case this giant magnon-like dispersion relation 
reduces to the same expression as that 
for a single one-spin giant magnon living in the $S^2$ subspace 
extracted from the giant magnon solution in the $SU(2)\times SU(2)$
 sector which consists of two giant magnons, one for each
$SU(2)$ with opposite momenta \cite{GHO}. 
There were investigations of the giant
magnon string solutions in various backgrounds \cite{KNP,WH}. 
The giant magnon-like expression in (\ref{sqe}) takes a similar form to
the dispersion relation in ref. \cite{KNP}, where the giant magnon 
solution in the near horizon geometry of the NS 5-brane background was
constructed to be expressed in terms of several charges.

For the spike solution case we produce the following relation
between the logarithmic divergent quantities such as the energy $E$,
the spin $J_{\phi}$ and the angle difference $\Delta\psi$   
\begin{eqnarray}
\kappa E - \nu J_{\phi} - \frac{\omega\al}{\beta}\frac{\sqrt{2\la}}{2}
\Delta\psi &=& 2\sqrt{2\la} \frac{\omega\al^2}{\beta^2}
\int_{\xi_+}^{\pi/4}d\xi \frac{\cos2\xi}{\sqrt{\cos^22\xi_+ 
- \cos^22\xi}\sin2\xi } \nonumber \\
 &=& \sqrt{2\la} \frac{\omega\al}{\beta}\tilde{\xi},
\label{ede}\end{eqnarray}
where $\tilde{\xi} = \pi/2 - 2\xi_+$ and the logarithmic divergences
are suitably canceled out. We use two relations (\ref{lin}) and
$\kappa^2 - \nu^2 = \omega^2\al^2/\beta^2$ to eliminate the parameters
and rewrite (\ref{ede}) as
\begin{equation}
\sqrt{E^2 - J_{\phi}^2} - \frac{1}{2}\sqrt{2\la}\Delta\psi = 
\sqrt{2\la}\tilde{\xi},
\label{sqj}\end{equation}
while the other spin $J_{\psi}$ is evaluated as
\begin{equation}
J_{\psi} = \sqrt{2\la}\sin\tilde{\xi}.
\end{equation}
The spike-like dispersion relation (\ref{sqj}) has a common 
energy-dependent term $\sqrt{E^2 - J_{\phi}^2}$,
compared to the giant magnon-like dispersion relation (\ref{sqe}).
It is noted that we need an appropriate factor 1/2 in front of 
$\sqrt{2\la}\Delta\phi_2$ and  $\sqrt{2\la}\Delta\psi$ to derive
the dispersion relations (\ref{hah}) and (\ref{sqj}).

\section{Conclusion}

Using the specific spherical coordinate 
representation of $CP^3$ such that
each foliating surface at constant $\xi$ is expressed by the U(1) bundle
over $S^2 \times S^2$ with coordinates $\psi$ and $(\theta_i, \phi_i) \;
i=1,2$ we have constructed two kinds of string states describing the
single giant magnon and single spike solutions with two angular
momenta moving in two different subspaces of $R_t \times CP^3$.

In one subspace of $CP^3$ characterized by one choice $\theta_1 =
\theta_2 = \pi/2, \psi = 0$ which satisfy the equations of motion for
$\theta_i$ and $\psi$, we have derived the dispersion relation of
the single giant magnon specified by the two spins $J_1$ and $J_2$
associated with the angular  $\phi_1$ and $\phi_2$ directions.
We have observed that the resulting dispersion relation turns out
to be the same expression as that of the single big giant magnon solution
with two spins \cite{IS} derived by the algebraic curve prescription,
which reduces to that of the giant magnon in the $RP^2$ subspace of 
$CP^3$ \cite{GGY} when the finite spin is turned off. 

In the other subspace of $CP^3$ characterized by $\theta_1 = \theta_2
= \pi/2, \phi_1 = \phi_2 \equiv \phi$ which give the complementary
solution, we have found some linear relations
between the conserved charges and the angle difference
 with suitable coefficients
from which we have determined the dispersion relations of the single
giant magnon and single spike solutions specified by the two spins
$J_{\psi}$ and $J_{\phi}$ associated with the angular 
$\psi$ and $\phi$ directions. 
When the spin $J_{\phi}$ is turned off the dispersion
relation of the giant magnon becomes the same expression as that of
the single one-spin giant magnon living in the $S^2$ subspace
of $CP^3$ \cite{GHO}.

\end{document}